# Title: AI can help scientists publish less

**Author: Gianfranco Bertone**

Affiliation: Gravitation Astroparticle Physics Amsterdam (GRAPPA), University of Amsterdam, 1098 XH Amsterdam, The Netherlands

*We can do more than defend science from a flood of AI-assisted papers. Used well, AI offers a historic opportunity to correct distortions in the publication system, help us publish fewer and better papers, and give scientists back the time to do their best work.*

The best outcome of AI in science would be to let us publish less, not more. That claim sounds paradoxical only because, for decades, volume has become our default proxy for progress. But AI has broken the symmetry it rested on. Producing a paper is collapsing in cost; evaluating one is not. Devising research plans, writing and running code, drafting manuscripts and polishing prose have become faster and cheaper[1,2]; reading, refereeing and judging remain as slow and as human as ever. Once that asymmetry enters a culture already shaped by publication pressure, the literature grows faster than communities can seriously read, review and absorb. AI forces a structural rethinking of how scientific work is organized, rewarded and recognized, while providing some of the means to carry it out. It can lift routine labour from scientists, strengthen the front end of evaluation, and return to scientists the time to decide what matters. Under the right incentives, this can help us publish fewer papers, and better ones.

## Negative epistemic value

Much of the current debate about AI in science has focused, understandably, on risk. Some warn against indiscriminate adoption[3]. Others argue that these tools expose older failures in scientific practice, standards and incentive structures[4]. Another strand asks what understanding should mean in AI-assisted science.[5] All of that is useful. But the central danger is simpler and more immediate. AI is entering a publication system already swollen by proliferation, marked by signs of declining disruptiveness[6,7], and under growing pressure at the level of review and evaluation.[8] Under those conditions, scientific papers can acquire *negative* epistemic value: not because they are wrong, but because the understanding they add no longer compensates for the time and attention they draw away from editors, referees, readers and colleagues trying to place them within what is already known.

Scientific articles now serve three functions at once. They communicate knowledge, they serve as tokens of career advancement,[9] and they circulate as products in an economic system that generates revenue, prestige and market power for publishers. Those functions do not pull in the same direction. Once papers and citations become central to hiring, promotion and funding, while publication itself also becomes a commercial object, proliferation acquires a force of its own. Science fills with large bubbles of urgent-but-not-important writing: work that is timely, legible to evaluators, easy to package and profitable to circulate.

A large number of incremental papers is not, in itself, a problem. Science has always advanced cumulatively. A paper can be incremental and still be careful, necessary and important: a careful reanalysis of systematic uncertainties, a sharper argument that closes a loophole, a replication that clarifies whether a result is robust, a null result that closes off an appealing but ultimately wrong hypothesis. Kuhn described mature research as "normal science", a cumulative practice of puzzle-solving within an accepted framework.[10] Most of what that looks like in practice is correction, refinement, better measurement and repeated testing: that is how the mountain of knowledge is built. A paper can also be highly original but technically flawed, or stylistically impeccable but epistemically irrelevant. Originality, quality, and significance are different dimensions of scientific work, and treating them as though they were interchangeable obfuscates the source of the problem.

The real difficulty begins when the cost of producing papers falls much faster than the cost of judging and absorbing them. The crisis in peer review already makes the imbalance visible. Journals struggle to find reviewers; editors work under severe strain; researchers face a literature that exceeds any realistic capacity for serious assimilation.[8] If AI is used badly, it will intensify exactly this asymmetry. What once required substantial time and effort, from generating ideas and performing calculations to synthesizing results and writing the paper, can now be done far more cheaply and quickly, while reading, refereeing and integrating remain slow, human and scarce. Under those conditions, the system will be pushed toward fragmentation and collapse. When scarce time and judgment are drained by papers whose contribution no longer justifies what they demand from everyone else, they contribute negatively to the collective production of knowledge, slowing and hampering the rise of the mountain.

## How AI can help

In Plato's *Phaedrus*, writing appears as a *pharmakon*: both a remedy and a poison. AI belongs squarely in that tradition. In the current publication culture it behaves as a poison, making it easier to produce claims than to test, review and place them. Used differently, it can become part of the remedy. I see major opportunities on three fronts, each mattering directly for how knowledge is created and evaluated.

The first concerns the visibility of non-paper contributions, such as code on GitHub, data on Zenodo, curated benchmarks, public notebooks, reproducibility packages, and living syntheses. They have existed for years and are valued by practising scientists. But they have remained second-class citizens: largely invisible to hiring and evaluation committees, and usually legitimized only via a paper that describes them rather than recognized in their own right. AI can change that communication layer. A codebase, dataset or benchmark need no longer remain intelligible only to insiders or only through the wrapper of a paper; it can become legible in its own right, through a living interface that shows what it does, how it is used, what depends on it, and what difference it has made. Much of what papers have long done for such work has been to make its value visible. AI can help do that job directly.

The second concerns time. AI can absorb much of the routine labour that now consumes researchers' effort: literature mapping, code scaffolding, documentation, reproducibility checks, exploratory work on alternative paths, first-pass synthesis across neighbouring literatures. That matters because it changes what scientific time can be used for. It can give researchers more room to pursue difficult questions, move into unfamiliar territory, test vague intuitions, abandon dead ends early, and follow genuine curiosity without every deviation immediately appearing as a publication deficit. The point is not that AI will tell scientists what is worth doing. It is that it can remove some of the weight that now pushes many worthwhile directions out of reach.

The third concerns evaluation. AI can strengthen the front end of review by helping editors, evaluation panels, and funders with triage, novelty checks, literature comparison, technical consistency checks and the detection of obvious pitfalls. None of this removes the need for human responsibility. Judgment remains human because significance remains human. What these tools can do is improve the informational basis on which judgment rests, while reducing some of the mechanical burden that now weighs on already overextended reviewers and panels. This is one of the places where AI can directly counter the danger of negative epistemic value. When scarce time is spent processing papers that add too little in return, knowledge suffers. Review would still take time where needed. What could change is the amount of low-level labour surrounding it, so that more of the community's limited attention is reserved for contributions that genuinely deserve it.

These changes point toward a different relation between scientific work and the paper. To publish less would not mean doing less science. It would mean recognizing more scientific contributions in their own right, rather than only after they have been converted into article form, and reducing the number of papers and applications whose attention cost exceeds their contribution to knowledge.

## What still has to be built

These possibilities will not realize themselves. If AI is going to mediate more of the labour of science, the scientific community cannot afford to become merely the customer of systems built by private companies whose incentives are only partly aligned with those of science. There is already an active debate on what a public scientific AI infrastructure should look like, with proposals ranging from CERN- or EMBL-style consortia to more distributed arrangements[11]. The specific institutional form is not the argument of this piece. What matters, whatever form emerges, is that the essentials are maintained as a public good: computing infrastructure under public control, transparent AI models that the community can host and run, open scientific knowledge databases, and dedicated tools to support review and evaluation.

Infrastructure, however, is only one side of the problem. Evaluation has already begun to move in the right direction: away from journal impact factors, toward narrative CVs, and toward a broader recognition of first-class scientific contributions[12]. But those measures were conceived for a slower publication economy. Once papers become cheaper to produce and easier to multiply, sustainability requires a criterion of responsible publication: not a cap on output, but a norm that judges publication against the attention it demands from the community. Fewer papers, when they reflect judgment rather than inactivity, should count as a mark of maturity and responsibility, not as a weaker claim on a field's attention.

Reducing the pressure to publish continuously would give scientists more time, and with it more freedom. Many researchers remain tethered to the logic of their doctoral work or to overcrowded lines of work, not because that is where their judgment leads them, but because changing direction implies lower output, diminished visibility and weaker prospects. Relieving that pressure would make it easier to pursue deep questions, move into a new field, learn a different method, or take a risk on a problem that may not yield quick returns.

This is also why philosophy of science should return to the working conversation of science and its evaluation. AI now makes urgent the questions that cannot be settled by paper counts or journal prestige. Is a field growing substantively, or only proliferating? Is a line of work deepening, or merely persisting? Is a research programme progressive or degenerative?[13] These questions are the business of philosophy of science, and they invite an exchange that is long overdue. That exchange cuts both ways: philosophers of science can see, from closer range, how AI is reshaping scientific practice, while scientists gain better tools for evaluating individual contributions and thinking together about what that practice is producing.

The opportunity opened by AI is therefore larger than the current debate often suggests. Used badly, it will flood the record with cheaper papers and push human evaluation beyond breaking point. Used well, it can help create a scientific culture in which more value is recognized before it is forced into article form, in which routine labour weighs less heavily on researchers, and in which significance and judgment return to the centre of our intellectual life where they belong. We should aim for more than defending science against a new technology. We can be bolder, and use it to make science, and the lives of scientists, better.

**Competing interests:** The author declares no competing interests.

**References**

1. Villaescusa-Navarro, F. et al. Preprint at https://arxiv.org/abs/2510.26887 (2025).

2. Lu, C. et al. Nature https://doi.org/10.1038/s41586-026-10265-5 (2026).

3. Trotta, R. Nat. Astron. 9, 1748–1749 (2025).

4. Peiris, H. V. Nat. Astron. (2026).

5. Ting, Y.-S., Curtis-Trudel, A. & Yao, S. Nat. Astron. https://doi.org/10.1038/s41550-026-02809-6 (2026).

6. Bhattacharya, J. & Packalen, M. NBER Working Paper 26752 (2020).

7. Park, M., Leahey, E. & Funk, R. J. Nature 613, 138–144 (2023).

8. Adam, D. Nature 644, 24–27 (2025).

9. Trueblood, J. S. et al. Proc. Natl Acad. Sci. USA 122, e2401231121 (2025).

10. Kuhn, T. S. The Structure of Scientific Revolutions 2nd edn (Univ. Chicago Press, 1970).

11. United Nations Secretary-General's AI Advisory Body. *Governing AI for Humanity: Interim Report* (United Nations, 2023). https://www.un.org/sites/un2.un.org/files/un_ai_advisory_body_governing_ai_for_humanity_interim_report.pdf

12. Coalition for Advancing Research Assessment. Agreement on Reforming Research Assessment (2022); https://www.coara.org/agreement/the-agreement-full-text/

13. Lakatos, I. The Methodology of Scientific Research Programmes (Cambridge Univ. Press, 1978).